\documentclass[letterpaper, 12pt]{article}

\makeatletter 
\renewcommand\section{\@startsection {section}{1}{\z@}%
                                    {-3.5ex \@plus -1ex \@minus -.2ex}%
                                    {0.5ex \@plus.2ex}%
                                    {\bf\scshape}} 
\renewcommand\subsection{\@startsection {subsection}{1}{\z@}%
                                    {-3.5ex \@plus -1ex \@minus -.2ex}%
                                    {0.5ex \@plus.2ex}%
                                    {\bf\small\scshape}}

\makeatletter

\usepackage{graphics,graphicx}
\usepackage{natbib}
\usepackage{setspace}
\usepackage{multicol} 
\usepackage{natbib}  
\usepackage{ulem}
\usepackage{color}
\usepackage{graphicx}
\usepackage{times}
\usepackage{amsmath}
\usepackage{amssymb}
\usepackage{url}
\usepackage[small,bf]{caption}
\usepackage{sidecap}
\usepackage{wrapfig}
\usepackage{xspace}
\usepackage{paralist}
\newif\ifAMStwofonts
\AMStwofontstrue
\usepackage{enumitem}
\usepackage{hyperref}
\usepackage{aas_macros}

\newcommand{\nustar}{\textsl{NuSTAR}\xspace}

\newcommand{\ergps}{\ensuremath{\text{erg\,s}^{-1}}\xspace}

\begin{document}
%
%
%
%

\begin{center}
\vspace*{4cm}    

\Huge{Breaking the limit: Super-Eddington accretion onto black holes and neutron stars}
\vspace{1cm}

\Large{Astro 2020 white paper}\\
\Large{Science area: {\bf Formation and evolution of compact objects}}

\vspace{1cm}

\Large{M. Brightman$^1$, M. Bachetti$^{2,1}$, H.~P. Earnshaw$^1$, F. F\"urst$^3$, J. Garc{\'\i}a$^{1,4}$, B. Grefenstette$^1$, M. Heida$^1$, E. Kara$^{5,6,7}$, K.~K.~Madsen$^1$, M.~J. Middleton$^8$, D. Stern$^9$, F. Tombesi$^{5,7,10,11}$, D.~J. Walton$^{12}$} 

\vspace{0.5cm}

\small{{\it $^1$ Space Radiation Laboratory, Caltech, 1200 E California Blvd, Pasadena, CA 91125; \\
$^2$ INAF-Osservatorio Astronomico di Cagliari, via della Scienza 5, I-09047 Selargius (CA), Italy\\
$^3$ European Space Astronomy Centre (ESAC), Science Operations Departement, Villanueva de la Ca\~nada, Madrid, Spain\\
$^4$ Dr. Karl-Remeis Observatory, Sternwartstr. 7, 96049 Bamberg, Germany\\
$^5$ Department of Astronomy, University of Maryland, College Park, MD 20742\\
$^6$ Joint Space Science Institute, University of Maryland, College Park, MD, 20742\\
$^7$ X-ray Astrophysics Laboratory, NASA/Goddard Space Flight Center, Greenbelt, MD 20771\\
$^8$ Department of Physics and Astronomy, University of Southampton, Highfield, Southampton SO17 1BJ, UK\\
$^9$ Jet Propulsion Laboratory, California Institute of Technology, 4800 Oa Grove Drive, Pasadena, CA 91109, USA\\
$^{10}$ Department of Physics, University of Rome ``Tor Vergata,'' Via della Ricerca Scientifica 1, I-00133 Rome, Italy\\
$^{11}$ INAF-Astronomical Observatory of Rome, Italy\\
$^{12}$ Institute of Astronomy, University of Cambridge, Madingley Road, Cambridge CB3 0HA, UK}}
\end{center}

\newpage

\section{Executive Summary}

With the recent discoveries of massive and highly luminous quasars at high redshifts ($z\sim7$; e.g. \citealt{mortlock2011}), the question of how black holes (BHs) grow in the early Universe has been cast in a new light. In order to grow BHs with $M_{\rm BH} > 10^9$\,M$_{\odot}$ by less than a billion years after the Big Bang, mass accretion onto the low-mass seed BHs needs to have been very rapid \citep{volonteri2005}. Indeed, for any stellar remnant seed, the rate required would need to exceed the Eddington limit. This is the point at which the outward force produced by radiation pressure is equal to the gravitational attraction experienced by the in-falling matter. In principle, this implies that there is a maximum luminosity an object of mass $M$ can emit; assuming spherical accretion and that the opacity is dominated by Thompson scattering, this Eddington luminosity is $L_{\rm{E}} = 1.38 \times 10^{38} (M/M_{\odot})$\,\ergps. In reality, it is known that this limit can be violated, 
due to non-spherical geometry or various kinds of instabilities.
Nevertheless, the Eddington limit remains an important reference point, and many of the details of how accretion proceeds above this limit remain unclear. Understanding how this so-called super-Eddington accretion occurs is of clear cosmological importance, since it potentially governs the growth of the first supermassive black holes (SMBHs) and the impact this growth would have had on their host galaxies (`feedback') and the epoch of reionization, as well as improving our understanding of accretion physics more generally.

\vspace{-0.2cm}
\section{Opportunities to Study Super-Eddington Accretion}

While in the early Universe the conditions for super-Eddington accretion onto black holes, such as gas density and metallicity, are thought to have been favourable, in the post-reionization Universe (which is far more readily observable) super-Eddington accretion onto SMBHs appears to be comparatively rare (e.g. \citealt{aird2018}). Nevertheless, there are rare examples of actively accreting SMBHs, otherwise known as active galactic nuclei (AGN), where the process super-Eddington accretion on to SMBHs can be observed directly (e.g. \citealt{lanzuisi2016,tsai2018}; Section \ref{sec_smbh}). However, since the emission from AGN accretion disks peaks in the ultraviolet (UV), absorption by neutral hydrogen often means that we cannot directly study emission from the hottest part of the accretion disk, where much of the gravitational potential energy is liberated. We therefore also need alternative means to improve our understanding of the super-Eddington regime.


Tidal disruption events (TDEs) occur when a star's orbit brings it close enough to a black hole such that tidal stresses overcome internal binding forces and it is torn apart, with a large fraction of its mass subsequently accreted at extreme rates. Although rare, these events are believed to be key laboratories for super-Eddington accretion \citep{rees1988}, and may be our best chance of directly studying highly super-Eddington accretion onto SMBHs (Section \ref{sec_tde}). However, this super-Eddington phase is only temporary, thus diminishing our opportunities to study it.


Ultraluminous X-ray sources (ULXs) are bright, persistent X-ray sources in galaxies that are not associated with their central SMBHs. They exhibit $L_{\rm{X}} > 10^{39}$\,\ergps\ which is the Eddington luminosity for a standard 10$M_{\odot}$ stellar remnant BH. Although once considered good intermediate-mass black hole (IMBH) candidates thanks to their extreme luminosities (e.g. \citealt{colbert1999}), our understanding of their spectral and timing properties now suggests that the majority of ULXs are stellar-mass compact objects undergoing super-Eddington accretion (e.g. \citealt{sutton2013,bachetti2013,walton2014,middleton2015,kitaki2017}; Figure 1 \textit{left panels}, and Section \ref{sec_ulx}). The Galactic source SS 433 is undoubtedly a  super-Eddington accreting object (see the review of \citealt{fabrika2004}), although it does not manifest as a ULX due to being aligned edge-on with respect to Earth \citep{begelman2006,middleton2018}. 

Since the basic physics of accretion do not depend on the mass of the accretor \citep[e.g.][]{mchardy2006,king2008,ruan2019}, by combining observations of relatively local super-Eddington SMBHs (AGN, TDEs) and X-ray binaries (ULXs) it will be possible to build up a coherent picture of this extreme regime, and substantially improve our understanding of black hole growth and feedback in the early universe which is extremely challenging to observe directly.





\vspace{-0.2cm}
\section{The State of the Art: Models and Simulations}

\begin{figure*}
\begin{center}
\hspace*{0.25cm}
\includegraphics[width=0.95\textwidth]{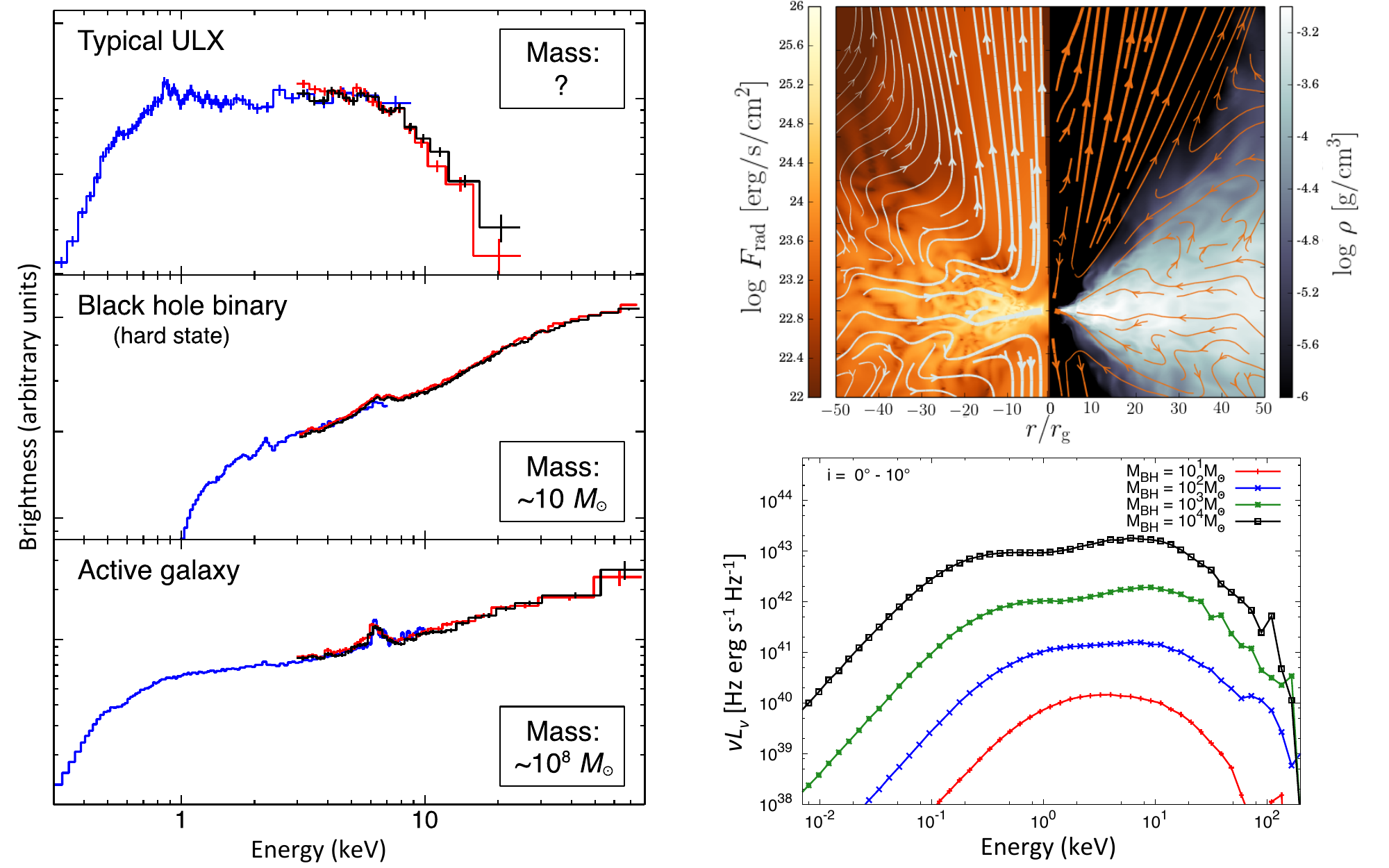}
\end{center}
\vspace{-0.75cm}
\caption{\footnotesize
\textit{Left panels:} example X-ray spectra of a typical ULX (\textit{top}), compared with a black hole X-ray binary in the hard state (\textit{middle}), and an accreting SMBH (\textit{bottom}). The ULX is clearly different to the other two systems, which are both accreting below the Eddington limit (and have similar X-ray spectra). This difference is especially evident above 10\,keV. \textit{Top right panel:} results from the numerical simulations of super-Eddington accretion presented by \citet{sadowski2016}. The left half shows the intensity of the radiative flux, with the arrows showing the streamlines followed by the radiation field (thicker lines mean higher flux). The right half shows the gas density, with the arrows showing the velocity of the gas (thicker lines mean higher velocity). The accretion flow clearly has a large scale-height, launches a strong disk wind, and preferentially funnels the radiation that escapes from the inner regions towards the poles.
\textit{Bottom right panel:} X-ray spectra predicted by similar simulations of super-Eddington accretion, assuming a variety of black hole masses (with $\dot{M} = 10^3 L_{\rm Edd}/c^2$ in all cases; from \citealt{kitaki2017}). The spectra are dominated by two blackbody-like components (with different relative contributions and temperatures), and qualitatively resemble the broadband X-ray spectra seen from ULXs.}
\label{fig_kitaki}
\vspace{-5mm}
\end{figure*}

In standard accretion disk theory, viscous heating in the disk is balanced by radiative cooling and the disk is geometrically thin and optically thick \citep{shakura1973}. However, at high mass accretion rates, the radiation pressure becomes substantial, increasing the disk scale-height ($H/R$). At the radius where the accretion rate reaches the local Eddington limit, the flow is said to be `super-critical' and $H/R$ tends to unity where gravity is balanced by radiation pressure. This results in the inner regions of the flow taking on a more funnel-like geometry, which can geometrically collimate the emission from the hottest regions of the flow, further enhancing the apparent luminosity (depending on the viewing angle). Mass must then be lost to prevent super-Eddington accretion rates at smaller radii where the radiative efficiency is higher \citep{shakura1973} and strong, radiatively driven disk winds are therefore universally predicted. In addition to this mass loss, advection is expected to provide a further means to cool the disk \citep{abramowicz1988}, and subsequently allow for even higher accretion rates. The combination of these two effects leads to our current best analytical description of such flows \citep{poutanen2007}.
Several works have conducted numerical simulations of super-Eddington flows (e.g. \citealt{ohsuga2009}; \citealt{ohsuga2011}; \citealt{jiang2014}; \citealt{sadowski2014}; \citealt{kitaki2017}), which show many of the characteristics outlined by the earlier works of \citet{shakura1973} and \citet{poutanen2007}; (see Figure \ref{fig_kitaki}, \textit{right panels}). The energy spectra predicted by these simulations are dominated by two multicolor blackbody components: one from the photosphere of the wind/outer disk, and one from the funnel-like inner regions of the flow (\citealt{kitaki2017}, \citealt{narayan2017}).

\section{The State of the Art: Observational Constraints}

\subsection{Super-Eddington accretion onto SMBHs}
\label{sec_smbh}


While interstellar hydrogen and atmospheric absorption complicate studies of the AGN accretion disk, it can be viewed indirectly through the study of X-rays reflected from it. Reflection features, such as fluorescent emission from iron and Compton-scattered photons, are imprinted on the intrinsic X-ray spectrum, and can reveal details about the geometry of the accretion flow (e.g. \citealt{ross2005, garcia2010, Wilkins12, Dauser16}), including the scale-height of the disk (\citealt{Taylor18}). This requires an X-ray telescope which is sensitive to energies around the iron K$\alpha$ line at 6.4 keV, and the peak of the Compton-scattered emission at 30 keV, which {\it NuSTAR} provides. {\it NuSTAR} observations of one of the brightest super-Eddington AGN, PG\,1247+267, revealed extreme reflection from the accretion disk in this object, indicative of a rapidly spinning BH \citep{lanzuisi2016}. 

\begin{wrapfigure}{R}{0.45\textwidth}
\begin{center}
\includegraphics[trim=10 10 10 30, width=75mm]{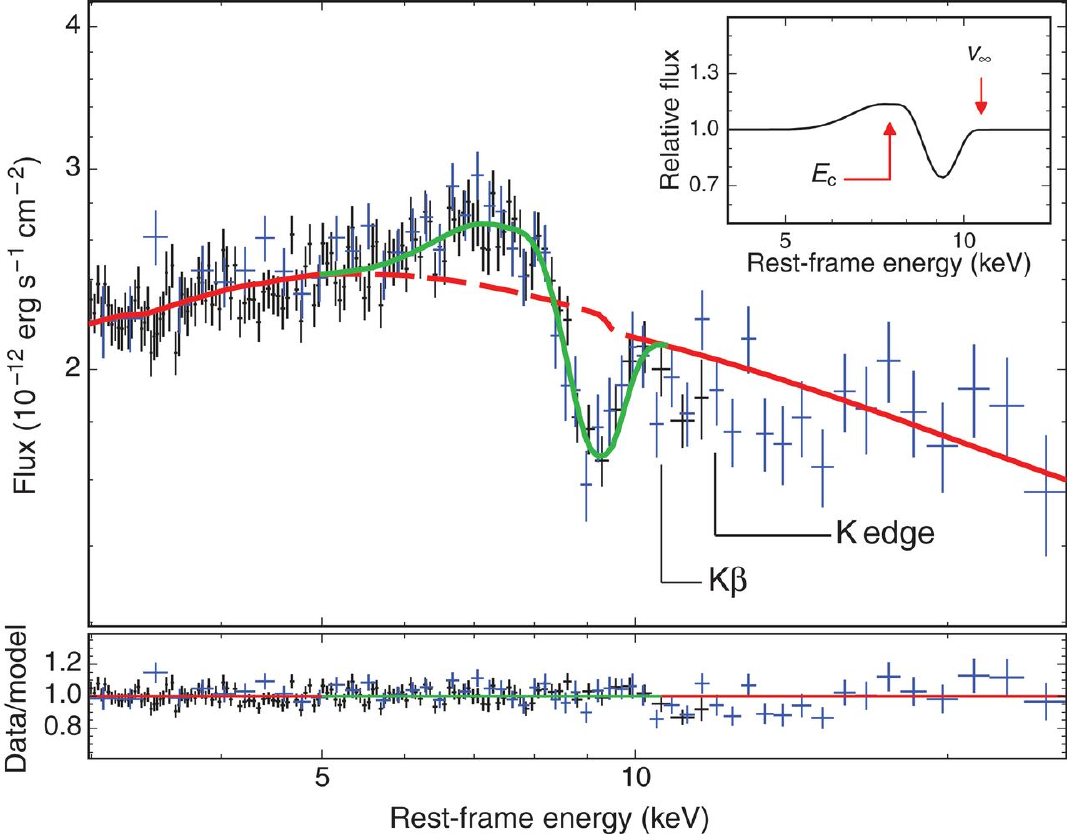}
\end{center}
\vspace{-0.25cm}
\caption{\footnotesize
Broadband X-ray spectrum of the luminous quasar PDS 456, taken from \citet{nardini2015} who used {\it XMM-Newton} and {\it NuSTAR} data. The spectrum shows a characteristic P-Cygni profile at 6--10 keV revealing the presence of high-velocity, outflowing, iron-rich gas as expected from high Eddington-rate systems. Only with sensitive observations above 10 keV from \nustar\ could the continuum shape be accurately determined such that the line profile could be elucidated.}
\label{fig_pds456}
\vspace{-0.5cm}
\end{wrapfigure}

Narrow Line Seyfert 1 galaxies are low-mass AGN that are also generally expected to be accreting at mildly super-Eddington rates. Recent studies of prominent NLS1s have revealed evidence for extreme disk winds with outflow velocities of $>$0.1$c$ (e.g. \citealt{parker2017}, \citealt{pinto2018}, \citealt{kosec2018}), as predicted by super-Eddington accretion models (e.g., \citealt{zubovas2012}. Similar winds have also been observed in a handful of AGN accreting close to the Eddington limit (\citealt{chartas2003}; \citealt{lanzuisi2012}; \citealt{vignali2015}; \citealt{tombesi2017}). However, fully characterising these winds requires broadband spectral coverage. A well known example of this is the luminous quasar PDS 456 (e.g. \citealt{reeves2003}. While data below 10 keV revealed blue-shifted absorption by extreme velocity outflowing iron, it was only the addition of data above 10 keV that allowed the characterization of the high-energy continuum which revealed the true line profile \citep{nardini2015}. This broadband data unveiled a broadened Fe K$\alpha$ emission component with an adjacent absorption trough, giving rise to a P-Cygni-like profile, as known to be produced by expanding gas (Figure \ref{fig_pds456}). This ultrafast outflow and others like it have kinetic energy with the potential to impart significant power into their host galaxies, thereby affecting their evolution. Therefore, understanding these outflows is not only important for studies of extreme accretion, but also to understand galaxy-SMBH co-evolution \citep{tombesi2015,feruglio2015}. We need to increase the number of sources available to study, and an instrument more sensitive than {\it NuSTAR} at energies above 10 keV is required.


\vspace{-0.2cm}
\subsection{Tidal disruption events}
\label{sec_tde}



TDEs are, by their nature, transient events that last for months to years. These events allow us to probe a population of otherwise quiescent SMBHs, and in some cases the fallback accretion rates from the disrupted star can exceed the Eddington rate. A well known example of this is the event Swift~J1644+57 (\citealt{bloom2011}; \citealt{burrows2011}) which reached a peak X-ray isotropic luminosity of $L_{\rm X}\sim10^{48}$ erg\,s$^{-1}$. The highly variable, broadband X-ray emission from this event allowed X-ray reverberation studies, which showed that it was powered by a BH of a few million solar masses, such that it reached 100 times its Eddington limit (although much of this was likely beamed; \citealt{kara2016}). The X-ray observations also indicated a large scale-height accretion flow as predicted by models of super-Eddington accretion. Evidence for extreme winds have also now been observed in a few systems (e.g. \citealt{lin2015}, \citealt{kara2018}), also consistent with super-Eddington predictions.

Swift~J1644+57 also highlights a current issue in the field. While fall back rates onto the SMBH are expected to be highly super-Eddington, from their optical emission they often appear to be sub-Eddington, creating a missing energy problem. Studying optically-selected TDEs in X-rays is therefore important for measuring their bolometric luminosity, and for understanding how much material is actually forming an accretion disk. While in the optical and UV some emission is expected to come from debris stream collisions, this is not the case for X-rays which are thought to originate from the newly formed disk.



The number and diversity of TDEs is set to explode due to upcoming wide-field surveys from the ground (e.g.~LSST) and space (e.g.~eROSITA). A sensitive, broadband X-ray observatory is clearly required in order to take advantage of this new era of discovery.

\vspace{-0.2cm}
\subsection{Ultraluminous X-ray Sources and Super-Eddington Neutron Stars}
\label{sec_ulx}

\begin{wrapfigure}{r}{0.4\textwidth}
\begin{center}
\vspace{-2\baselineskip}
\includegraphics[width=70mm]{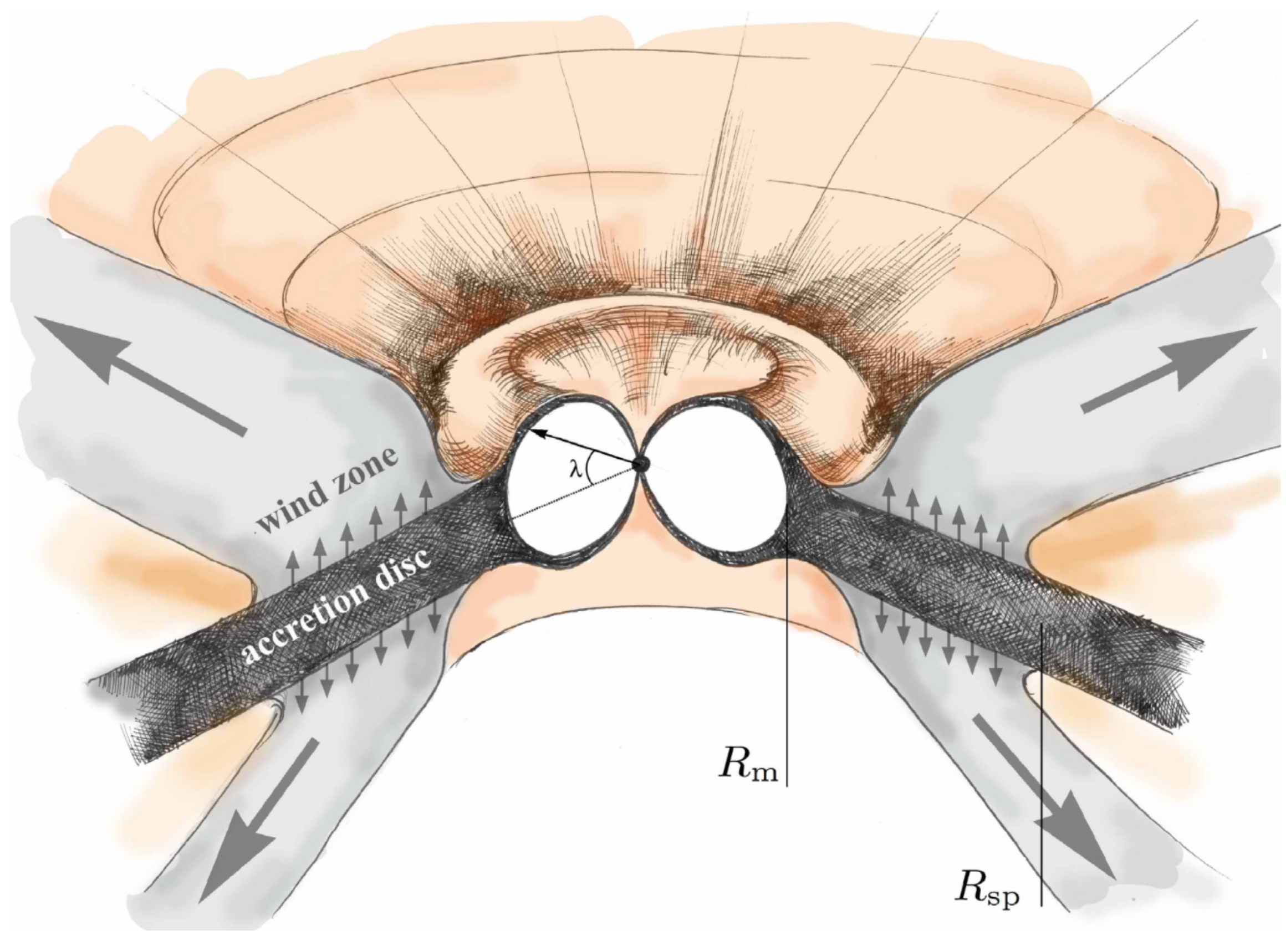}
\end{center}
\vspace{-0.75cm}
\caption{\footnotesize
Artistic impression of a ULX powered by accretion on to a strongly magnetized neutron star, taken from Mushtukov et al (2019). The accretion disk is truncated from the inside at the magnetospheric radius, $R_{\rm m}$, due to interaction with the magnetic field of the NS, which funnels the accreted material onto the magnetic poles. }
\label{fig_mushtukov}
\end{wrapfigure}




Through the discovery of pulsations, \citet{bachetti2014} showed, remarkably, that the accretor in the ULX M82 X-2 is a neutron star (NS). Because this source is known to reach X-ray luminosities of $\sim 2 \times 10^{40}$ erg\,s$^{-1}$, and NSs only occupy a small mass range (1--2 M$_{\odot}$), this discovery implied that it was exceeding its Eddington limit by a factor of $\sim100$!
Since then, several other NS ULXs have been discovered, one of which, NGC~5907~ULX1, has an apparent luminosity around 500 times the Eddington limit (\citealt{israel2017sci}, \citealt{furst2017}), challenging current accretion models. 
In our own Galaxy, the transient NS Swift~J0243.6+6124 can reach luminosities of $\sim$2 $\times 10^{39}$\,\ergps\ ($\sim$10 $L_{\rm{E}}$), making it the brightest known NS in the Milky Way and an ideal object to study the transition to the super-Eddington accretion regime \citep{wilsonhodge2018}.
Because NS ULXs are the best evidence for sustained super-Eddington accretion, discovering more of these sources and understanding their relative contribution to the overall ULX population is one of the main challenges for the next decade.

The X-ray spectrum can also contain telltale signs of a NS accretor. For example, cyclotron resonance scattering features (CRSFs), produced by charged particles in the presence of the neutron star's strong magnetic field.
Such a feature has been found in M51 ULX8 \citep{brightman2018} and implied in another \citep{walton2018crsf}, already known to be a NS. 
Detecting and studying such features in other ULXs has the potential to reveal information about the magnetic field strength and geometry, perhaps elucidating how these neutron stars can accrete at such extreme rates.

Although the pulsations clearly require magnetically channelled accretion (see Fig. \ref{fig_mushtukov}), the broadband spectral shape of the known NS ULXs may be significantly influenced by the super-Eddington accretion flow beyond the magnetosphere \citep[e.g.][]{walton2018mnras}. Such flows should be similar for both BHs and NSs \citep{king2008}. Indeed, qualitatively similar X-ray spectra are found for {\it all} ULXs with broadband (e.g. \textit{XMM-Newton}+\textit{NuSTAR}) observations to date \citep[see also Fig. 1]{koliopanos2017,pintore2017,walton2018mnras,walton2018apj}, implying that either the population is dominated by neutron star accretors, or that the accretion flows in NS and BH ULXs (if the latter are present in the sample) really are similar.  
In addition, the ultrafast outflows that are thought to be a hallmark of super-Eddington accretion have recently been detected in several ULXs, including one where the accretor is know to be a neutron star \citep{pinto2016,walton2016,kosec2018}.

However, the current sample of well-studied ULX is relatively small ($\sim$15 sources).
More sensitive instruments for broadband X-ray spectroscopy, combined with excellent X-ray timing capabilities, will greatly increase the number of sources where we can detect pulsations, cyclotron resonance features, and outflows. A focusing X-ray telescope capable of observing at energies above 10\,keV, more sensitive than {\it NuSTAR}, is especially important, as that is the regime where the difference between sub- and super-Eddington spectra is most pronounced (see Fig.~\ref{fig_kitaki}). With a larger sample of ULXs with high-quality broadband X-ray spectra, we can search for features unique to black hole accretors, determine the effects of inclination and accretion disk precession, and investigate how super-Eddington accretion in ULXs relates to the growth of SMBHs in the early Universe. Finally, it should be noted that there exists no ULX where the presence of a black hole accretor is irrefutable, which is a major step needed to understand super-Eddington accretion.



\vspace{-0.2cm}
\section{Outstanding Questions}

Our current X-ray facilities have enabled us to make excellent progress towards understanding the extremes of accretion onto compact objects. However, there are still a number of outstanding questions and limitations on our current understanding that will require the next generation of instruments to answer:

\vspace{0.2cm}
\begin{itemize}[noitemsep,topsep=0pt,parsep=0pt,partopsep=0pt]
    \item {\bf Is there evidence that black holes undergo persistent super-Eddington accretion?} 
    We currently have no direct evidence that an individual black hole undergoes the persistent super-Eddington accretion that is required for SMBHs to acquire their masses by $z\sim7$.  
    Increasing the number of observable persistently accreting black holes is key to evaluate the viability of this mechanism.
    \item {\bf How does the presence of a magnetic field affect super-Eddington accretion
    ?} It is still unclear what proportion of the population of ULXs are neutron stars, and how they relate to the progenitors of SMBHs. In order to answer this question we need the capability to conduct pulsation searches, identify cyclotron lines, and investigate new ways of identifying the nature of the compact objects.
    \item {\bf What level of accretion makes it onto the compact object, and how can sufficient mass delivery for super-Eddington accretion be achieved?} 
    The progenitors of SMBHs may have required sustained growth to reach the sizes required by our high-redshift observations of quasars. 
    Understanding how fast a compact object is accreting requires knowing both how much matter is fed onto the system, and how much is driven out in outflows. It is necessary to refine our understanding of the radiative efficiency of these systems, as well as to monitor the changes in super-Eddington accreting sources over time, and study transient systems such as TDEs.
    \item {\bf How does super-Eddington accretion affect the environment of a source and how is this related to feedback in galaxies?} Super-Eddington accreting sources may have played a role in the reionisation of the Universe and in the evolution of their host galaxies. Understanding this requires us to study the physics of outflows and their effect on their environment observable in bubbles and nebula surrounding the source.
\end{itemize}

\vspace{-0.2cm}
\section{Instrumental requirements}
\begin{itemize}[noitemsep,topsep=0pt,parsep=0pt,partopsep=0pt]
    \item {\bf A high-throughput (e.g. 5000\,cm$^{-2}$ at 10\,keV) and broadband (extending to \textit{E} $=$ 100\,keV) X-ray spectrometer.} This is necessary to fully characterize the emission from super-Eddington flows and the reflection features from AGN/TDE accretion disks, both disentangling the various emission components and providing tight constraints on key parameters. This will yield an order-of-magnitude improvement in sensitivity over {\it NuSTAR}, significantly increasing sample sizes. 
    \item {\bf Relative timing resolution on the micro-second level with a stable clock and fast readout.} This will allow the detection of pulsations required to identify neutron star accretors.
    \item {\bf Spectral resolution of 650 or better over the 0.3--30\,keV range.}
    This will allow us to investigate outflowing winds by observing their highly blueshifted atomic absorption. It will also allow us to identify narrow absorption lines that could be produced by proton CRSFs.
    \item {\bf Spatial resolution of 15'' or better.} This will allow us to resolve sources in relatively crowded fields of galaxies and attain good sensitivity by keeping backgrounds low.
    
\end{itemize}

\bibliographystyle{aa_url}
\bibliography{bibliography}

\end{document}